\documentclass[aps,pra,preprint,groupedaddress]{revtex4-1}

\usepackage{graphicx,graphics}
\usepackage{bm}
\usepackage{amsmath}
\begin{document}


\title{Inclusion of the backaction term in the total optical force exerted upon Rayleigh particles in nonresonant structures}


\author{Mohammad Ali Abbassi}
\email[]{mohammadali.abbassi@ee.sharif.edu}
\affiliation{Sharif University of Technology, Tehran, Iran}
\author{Khashayar Mehrany}
\email[]{mehrany@sharif.edu}
\affiliation{Sharif University of Technology, Tehran, Iran}

\date{\today}

\begin{abstract}
In this paper, we investigate the impact of the electromagnetic scattering caused by other objects in nonfree space on the time averaged force exerted upon a Rayleigh particle, which is conventionally referred to as the backaction effect. We show that backaction modifies the gradient force, radiation pressure, and spin curl force exerted upon Rayleigh particles, and gives rise to an additional force term which stems from the gradient of the backaction field in nonfree space. As a numerical example, we look into the trapping of a dielectric nanoparticle at the center of curvature of a spherical mirror, and study how it is affected by the backaction effect. We show that backaction can enhance the force exerted upon the particle, reshape the trapping potential, and shift the equilibrium position of the particle. 
\end{abstract}


\maketitle

\section{Introduction}
In the past decades, many impressive advances have been made in the field of optical manipulation of mesoscopic objects\cite{marago2013optical,bendix2014optical, gao2017optical}. The forces induced by the interaction of matter with the optical fields have been employed to realize optical tweezers\cite{marago2013optical,bendix2014optical, gao2017optical,moffitt2008recent,fazal2011optical,zhao2016enantioselective}, rotators\cite{li2016optically,yan2013optical,miyakawa2004rotation,la2004optical}, and tractor beams\cite{novitsky2012material,sukhov2010concept,ruffner2012optical,novitsky2011single,kajorndejnukul2013linear, brzobohaty2013experimental}. Despite the substantial progress made in the trapping of micrometer-sized objects, stable trapping of Rayleigh nanoparticles is still challenging in the free space because the polarizability of the particle decreases rapidly when its size is reduced\cite{gao2017optical,ashkin1986observation,chaumet2000time,albaladejo2009scattering}. The seemingly straightforward solution of increasing the intensity of the incident light to counterbalance the reduction of the polarizability is not viable since increasing the intensity of the light can damage the particle before it gets trapped\cite{grigorenko2008nanometric}. One possible approach to address this issue is to employ nanostructures which can confine the electromagnetic field beyond the diffraction limit\cite{grigorenko2008nanometric,juan2011plasmon,yang2009forces,yang2009optical, saleh2012toward,yang2011optical,jazayeri2017all,woolf2009forces,volpe2006surface,wang2009propulsion,righini2007parallel, righini2008surface}. In this fashion, the trapping force is enhanced by increasing the gradient of the incident field in the absence of the particle. Interestingly, there is a further enhancement of the trapping force reported in resonant nanostructures on account of the so-called self-induced backaction (SIBA) effect\cite{juan2009self,chen2011enhanced,pang2011optical,berthelot2014three,descharmes2013observation,mestres2016unraveling,luis2016arrested}. These structures are designed in such a way that the fulfillment of the resonance condition in the structure has a strong dependence on the position of the particle. In this fashion, the particle has a strong influence on the local electric field, and plays an active role in its trapping\cite{juan2009self}.\par
The conventional approach to studying the SIBA phenomenon observed in resonant structures is to employ the Hamiltonian formalism, whereby it can be easily shown that the SIBA can reshape the trapping potential, and can provide dark field trapping that reduces the averaged intensity seen by the particle \cite{neumeier2015self}. It should be noted, however, that applying the Hamiltonian formalism as presented in Ref. \cite{neumeier2015self} is solely valid when the structure has a high quality resonance. As a matter of fact, the quality factor of the resonance should be high enough to justify approximation of the electromagnetic fields by a single resonance mode of the structure. In other words, the impact of the continuum radiation modes of the structure should be negligible. That is why the conventional Hamiltonian formalism is not capable of analyzing the backaction effect in structures which do not have a high enough quality factor. However, the SIBA is not solely observed in resonant structures with high quality factors and can be present in structures either with low quality resonance or even no resonance at all. Moreover, the SIBA might bring about results other than the increased trapping force. For instance, long-range optical pulling force has been reported recently in a photonic crystal structure\cite{zhu2018self}. \par 
In this paper, we investigate the time averaged optical force exerted on Rayleigh particles in the vicinity of scattering objects which does not necessarily form a resonant structure. We formulate the backaction effect using the dyadic Green function of the structure, and demonstrate that it modifies the polarizability of the particle, and the total time-averaged force. We show that the polarizability of the particle is no longer a scalar value even for spherical particles, and becomes a dyadic quantity which depends on the position of the particle. Therefore, the gradient force, radiation pressure, and spin curl force are all modified on account of the backaction effect. Furthermore, an additional term, which is proportional to the gradient of the scattering Green function of the structure, appears in the time-averaged total force. It should be noted that defining the effective polarizability based on the Green function has already been reported in literature\cite{wang2016strong, petrov2016surface, novotny2012principles, chaumet2000coupled, van2011optical}. The main contribution of this paper is to show how the backaction field modifies different force terms, and how the new force term associated with the gradient of the scattering Green function could be the significant factor in determining the overall optical force. The proposed formulation remains exact insofar as the effects of higher multipole orders remain negligible. This is the matter that we have looked into numerically by comparing the results of our formulation against those obtained by applying the Maxwell stress tensor (MST). In this fashion, the significance of the higher multipole orders is quantitatively studied when Rayleigh particles come close to the boundaries. We also show that even though the proposed formalism is inherently different from the Hamiltonian formalism, both approaches provide the same results when the Green function in our proposed formulation is approximated by the electromagnetic fields of a single resonance mode. \par
The organization of this paper is as follows: In Sec. \ref{sec2}, the mathematical formalism for the inclusion of the backaction effect in the force exerted upon Rayleigh particles in nonfree space is presented. It is also shown that the proposed formulation yields essentially the same results that the more conventional Hamiltonian formalism provides. Two different scenarios are then studied numerically in Sec. \ref{sec3}. In Sec. \ref{sec3:a}, the backaction effect is investigated at the center of curvature of a spherical mirror, which is obviously a nonresonant structure. It is shown that the backaction can reshape the trapping force and potential, and also shift the equilibrium position of the particle. In Sec. \ref{sec3:b}, the trapping force exerted upon a Rayleigh nanosphere is studied in a structure with a circular nano-hole (CNH). The contribution of different force terms in the total optical force is looked into and it is numerically shown that the backaction effect is mainly the result of the scattering Green function gradient. In other words, the difference between the results of the Maxwell stress tensor method and the perturbative method reported in Ref. \cite{juan2009self} is not entirely due to the contribution of the higher order multipoles. Rather, the difference between the two methodologies is mostly attributable to the gradient of the scattering Green function which was hitherto neglected in the perturbative method. Eventually, the conclusions are made in Sec. \ref{sec4}.\par

\section{\label{sec2}Formulation}
In the Rayleigh regime where the size of the particle is much smaller than the wavelength of the electromagnetic field, the particle behaves as if it is an electric dipole\cite{ashkin1986observation,chaumet2000time,albaladejo2009scattering}, whose electric dipole moment is $\bm{p}=V_p\bm{P}(\bm{r}_p)$, where $V_p$ is the volume of the particle, and $\bm{P}(\bm{r}_p)$ is the polarization at the center of mass of the particle:
\begin{equation}\bm{P}(\bm{r}_p)=\epsilon_0(\epsilon_p-1)\left[\bm{E}_0(\bm{r}_p)+\bm{E}_p(\bm{r}_p)\right]\end{equation}
Here, $\epsilon_p$ is the relative permittivity of the particle, $\bm{E}_0$ is the electric field  in the absence of the particle, and $\bm{E}_p$ is the electric field generated in the presence of the particle. The latter can be written as a summation of two terms. One is the electric field radiated by the particle in the free space which does not contain the impact of other scattering objects, and is given by\cite{novotny2012principles}
\begin{equation}\label{eq:Erad}\bm{E}_{\mathrm{rad}}(\bm{r})=\left[\left(k_0^2\tensor{\bm{I}}+\bm{\nabla\nabla}\right)\int_{V_p}\frac{e^{ik_0|\bm{r}-\bm{r}'|}}{4\pi\epsilon_0|\bm{r}-\bm{r}'|}d^3\bm{r}'\right].\bm{P}(\bm{r}_p)\end{equation}
The other which is hereafter referred to as the backaction field is in fact the scattered field of the particle radiation, and can be written as
\begin{equation}\bm{E}_{\mathrm{ba}}(\bm{r})=V_p\tensor{\bm{G}}_s(\bm{r},\bm{r}_p).\bm{P}(\bm{r}_p)\end{equation}
where $\tensor{\bm{G}}_s$ is the scattering Green function of the structure, and is easily obtained by the elimination of the free space Green function from the total Green function of the structure.\par
It should be noted that $\bm{\nabla}$ operators and the integral in Eq. (\ref{eq:Erad}) cannot be interchanged since the result will be singular at $\bm{r}=\bm{r}'$\cite{lee1980singularity,yaghjian1980electric}. Since this singularity stems from the static part of $\bm{E}_{\mathrm{rad}}$, it can be easily dealt with once the static and dynamic contributions within the radiation field are separated from each other. Therefore, for further simplification, $\bm{E}_{\mathrm{rad}}$ can be written as
\begin{equation}\bm{E}_{\mathrm{rad}}=\bm{E}_{\mathrm{sta}}+\bm{E}_{\mathrm{dyn}}\end{equation}
The first term, $\bm{E}_{\mathrm{sta}}$, is the static part of the radiated field which is given by
\begin{equation}\bm{E}_{\mathrm{sta}}(\bm{r})=\tensor{\mathcal{S}}(\bm{r}) .\bm{P}(\bm{r}_p)\end{equation}
where
\begin{equation}\ \mathcal{S}_{ij}(\bm{r})=\partial_i\partial_j\int_{V_p}\frac{1}{4\pi\epsilon_0|\bm{r}-\bm{r}'|}d^3\bm{r}'=\oint_{S_p}\frac{x_i-x'_i}{4\pi\epsilon_0|\bm{r}-\bm{r}'|^3}\hat{x}_j.d\bm{s}'\end{equation}
The next term, $\bm{E}_{dyn}$ is the dynamic part of the radiated field which can be written as
\begin{equation}\bm{E}_{\mathrm{dyn}}(\bm{r})=\tensor{\mathcal{R}}(\bm{r}).\bm{P}(\bm{r}_p)\end{equation}
where
\begin{equation}\label{eq:R1}\tensor{\mathcal{R}}(\bm{r})=\frac{1}{4\pi\epsilon_0}\left[k_0^2\tensor{\bm{I}}\int_{V_p}\frac{e^{ik_0|\bm{r}-\bm{r}'|}}{|\bm{r}-\bm{r}'|}d^3\bm{r}'+\int_{V_p}\bm{\nabla\nabla}\frac{e^{ik_0|\bm{r}-\bm{r}'|}-1}{|\bm{r}-\bm{r}'|}d^3\bm{r}'\right]\end{equation}
and can be further simplified to
\begin{equation}\label{eq:R2}\tensor{\mathcal{R}}(\bm{r})=\frac{k_0^2}{8\pi\epsilon_0}\int_{V_p}\frac{|\bm{r}-\bm{r}'|^2\tensor{\bm{I}}-(\bm{r}-\bm{r'})(\bm{r}-\bm{r'})}{|\bm{r}-\bm{r}'|^3}d^3\bm{r}'+\frac{ik_0^3}{6\pi\epsilon_0}V_p\tensor{\bm{I}}\end{equation}
by expanding Eq. (\ref{eq:R1}) with respect to $k_0$ up to the third order term. It should be noted that while the second order term in the above expression can be neglected, keeping the third order term is necessary since it is the first imaginary term presents in the expansion. This term is usually referred to as the radiation reaction of the particle, which makes the polarizability a complex value even in the absence of absorption\cite{chaumet2000time,draine1988discrete}.\par
Now, it can be easily shown that the equivalent electric dipole moment is given by
\begin{equation}\bm{p}=\tensor{\alpha}.\bm{E}_0(\bm{r}_p)\end{equation}
Here, $\tensor{\alpha}$ is the dyadic polarizability of the particle and reads as
\begin{equation}\tensor{\alpha}=\left[\tensor{\bm{I}}-\frac{ik_0^3}{6\pi\epsilon_0}\tensor{\alpha}_0-\tensor{\alpha}_0.\tensor{\bm{G}}_s(\bm{r}_p,\bm{r}_p)\right]^{-1}.\tensor{\alpha}_0\end{equation}
where 
\begin{equation}\tensor{\alpha}_0=\epsilon_0\left(\epsilon_p-1\right)V_p \left[1-\epsilon_0(\epsilon_p-1)\tensor{\mathcal{S}}(\bm{r}_p)\right]^{-1}\end{equation}
is the static polarizability of the particle in the free space. It should be noted that for a spherical particle, $\tensor{\alpha}_0$ is isotropic and is given by
\begin{equation}\alpha_0=\epsilon_0V_p\frac{\epsilon_p-1}{\epsilon_p+2}\end{equation}
After finding the equivalent localized electric dipole, the time-averaged force exerted upon the particle can be obtained from
\begin{equation}\langle F \rangle= \frac{1}{2}\Re\left[\sum_i p_i \bm{\nabla}{E_{i}^\ast}\right]\end{equation}
where
\begin{equation}\label{eq:E_sep}\bm{E}=\bm{E}_{0}+\tensor{\bm{G}}_s (\bm{r}_p,\bm{r}_p).\bm{p}\end{equation}
is the electric field experienced by the electric dipole that consists of the electric field in the absence of the particle, $\bm{E}_{0}$ , and the backaction field due to the presence of scattering objects in the vicinity of the particle.
It is quite obvious that the time averaged total force can be separated to four terms:
\begin{multline}\label{eq:F2}\langle \bm{F}\rangle=\frac{1}{4} \sum_{ij} \alpha_{ij}^\prime \bm{\nabla} \left (E_{0_i}E_{0_j}^\ast \right) +\frac{\omega\mu_0}{2}\Re\left[\left(\tensor{\bm{\alpha}}^{\prime\prime}.\bm{E}_{0}\right)\times \bm{H}_{0}^\ast\right]\\+\frac{1}{2}\Im\left[\left(\left(\tensor{\bm{\alpha}}^{\prime\prime}.\bm{E}_{0}\right).\bm{\nabla}\right)\bm{E}_{0}^\ast\right]+\frac{1}{2}\Re\left[\sum_{ijkl}\alpha_{ik}\alpha_{jl}^\ast E_{0_k}E_{0_l}^\ast\bm{\nabla}G_{s_{ij}}^\ast\right]\end{multline}
where $\tensor{\alpha}'$ and $\tensor{\alpha}''$ are the real and imaginary parts of the polarizability, respectively.
The first term is the generalized form of the gradient force which clearly becomes the conventional form of the gradient force whenever $\tensor{\bm{\alpha}}'$ is isotropic\cite{albaladejo2009scattering,marago2013optical}. Moreover, the second and the third terms that depend on the imaginary part of the polarizability are the generalized forms of radiation pressure, and spin curl force, respectively. Eventually, the last term is a new force term that does not appear in the free space, and depends on the gradient of the scattering Green function.\par
As already mentioned, the proposed formalism is capable of modeling the backaction effect in both resonant and nonresonant structures. Here, we want to apply this formalism to investigate trapping of a Rayleigh nanosphere in a resonant structure, and show that it leads to the same results with the Hamiltonian formalism discussed in the appendix. In a resonant structure, the electromagnetic fields can be almost approximated with the resonant mode of the structure, and the impact of other modes can be neglected. It is worth noting that this approximation is only valid in high-Q resonant structures and near its resonance frequency. In such a case,
\begin{equation}G_s(\bm{r},\bm{r}_p)=\frac{-i\omega_L}{2\epsilon_0V_m}\frac{u(\bm{r})u^\ast(\bm{r}_p)}{(-\frac{\kappa}{2}-i\omega_c)+i\omega_L}\end{equation}
using the modal expansion of the Green function where the term corresponding to the resonant mode is just kept in the expansion. Here, $\omega_c-i\frac{\kappa}{2}$ is the complex resonant frequency of the structure, and $\omega_L$ is the frequency of the driving laser. Moreover, $u(\bm{r})$ and $V_m$ are the normalized field profile and mode volume of the resonant mode, respectively. Hence,
\begin{equation}G_s(\bm{r}_p,\bm{r}_p)=\frac{iA}{\alpha_0\left(\frac{\kappa}{2}-i\Delta\right)}\Psi(\bm{r}_p)\end{equation}
where $A=\frac{\omega_c\alpha_0}{2\epsilon_0V_m}$, and $\Psi(\bm{r}_p)=|u(\bm{r_p})|^2$ is the normalized intensity profile of the resonant mode. Moreover, $\Delta=\omega_L-\omega_c$ is the detuning from the resonance frequency of the structure. Now, the equivalent dipole moment can be written as
\begin{equation}p=\alpha_0 E(\bm{r}_p)=\frac{\alpha_0}{1-\alpha_0 G_s(\bm{r}_p,\bm{r}_p)}E_0(\bm{r}_p)=\frac{\alpha_0}{1-\frac{iA}{\frac{\kappa}{2}-i\Delta}\Psi(\bm{r}_p)}E_0(\bm{r}_p)\end{equation}
and thereby the effective polarizability of the nanosphere can be obtained from
\begin{equation}\alpha=\frac{\alpha_0}{1-\frac{iA}{\frac{\kappa}{2}-i\Delta}\Psi(\bm{r}_p)}\end{equation}
which is the same as the result of Hamiltonian formalism discussed in the Appendix.
According to the above equation, we can define the backaction parameter as $\eta=\frac{2A}{\kappa}$\cite{neumeier2015self}. When $\eta\ll 1$, the impact of the cavity on the polarizability of the nanosphere is negligible, and $\alpha\simeq \alpha_0$. However, for large $\eta$, the cavity can significantly change the polarizability of the nanosphere due to the backaction effect. Furthermore, when $\eta\gg 1$, $\alpha\simeq \frac{i\alpha_0}{\eta\Psi(\bm{r}_p)}$ for $\Delta=0$, which means that backaction makes the polarizability of the nanosphere become almost imaginary.
\section{\label{sec3}Numerical examples}
In this section, we present two numerical examples to observe the impact of backaction on the forced exerted upon a Rayleigh particle. In the first example, we investigate trapping of a Rayleigh nanosphere at the center of curvature of a spherical mirror. In this configuration, the spherical mirror can concentrate the backaction field at the position of the particle and thereby can affect the force exerted upon the particle. This example, which clearly has no resonance, demonstrates that backaction can also be observed in a nonresonant structure. Furthermore, it indicates that backaction can be observed in the far zone of a scattering object which to the best of our knowledge has not been reported yet.\par
 Then, as the second example, we investigate the trapping of a Rayleigh nanosphere in the circular nano-hole structure which counts among the earliest cases in which SIBA had been reported\cite{juan2009self}. We compare the results obtained from the proposed formalism with those calculated by the Maxwell stress tensor method, and show that the failure of the perturbative method reported in Ref. \cite{juan2009self} is not totally due to the failure of the Rayleigh approximation, i.e., the contribution of higher-order multipoles. It is shown that the dominant effect is the impact of the scattering Green function which was not considered in Ref. \cite{juan2009self}, especially when the particle is far from the walls of the hole. This is interpreted as the backaction effect in our paper.
\subsection{\label{sec3:a}Spherical mirror}
\begin{figure}
\includegraphics[width=0.4\textwidth]{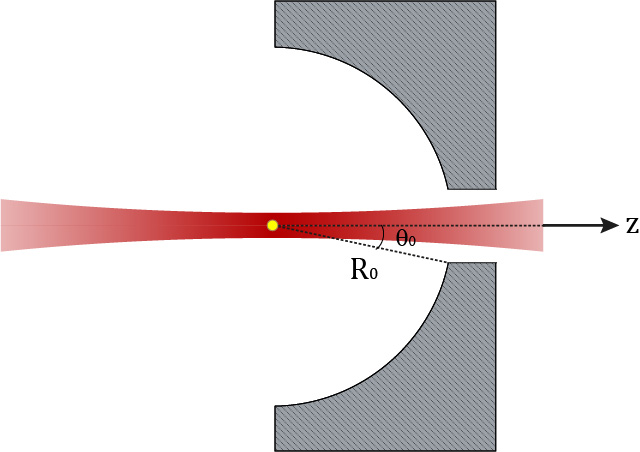}
\caption{\label{scheme}Schematic of the system. A dielectric nanosphere is trapped by an optical tweezer at the center of a spherical mirror}
\end{figure}
In this section, the trapping of a Rayleigh nanosphere at the center of curvature of a spherical mirror which focuses the backaction field at the position of the particle is studied to demonstrate the backaction effect in a nonresonant structure. The schematic of this system is depicted in Fig.\ref{scheme}. As shown in this figure, a dielectric nanosphere is trapped by a Gaussian beam which passes through a spherical mirror with radius $R_0$ stretched from $\theta_0$ to $\frac{\pi}{2}$ around the $z$ axis. It is worth noting that the Gaussian beam propagates along the $z$-axis, has free space wavelength $\lambda_0$ and its focal point is adjusted at the center of the spherical mirror.\par
The electric field in the absence of the nanoparticle in the proposed structure is given by
\begin{equation}\bm{E}_0(\bm{r})=\sqrt{2\eta_0I_0}\hat{x}\frac{w_0}{w(z)}\exp\left(-\frac{x^2+y^2}{w^2(z)}\right)\exp\left(i\Big(k_0z+k_0\frac{x^2+y^2}{2\mathcal{R}(z)}-\psi(z)\Big)\right)\end{equation}
where $I_0$ is the intensity of the beam at its waist, and $w_0$ is the beam waist radius. Furthermore, $w(z)=w_0\sqrt{1+\frac{z^2}{z_R^2}}$, and $\mathcal{R}(z)=z(1+\frac{z_R^2}{z^2})$ are the beam radius, and wavefront radius, respectively. Eventually, $\psi(z)=\arctan(\frac{z}{z_R})$ denotes the Gouy phase of the optical beam, and $z_R=\frac{\pi w_0^2}{\lambda_0}$ is called the Rayleigh distance from the beam waist\cite{novotny2012principles}.\par
Since the radius of the spherical mirror is much larger than the wavelength of the optical beam ($R_0\ll\lambda_0$), we can use physical optics approximation to calculate the dyadic scattering Green function of the structure which reads as
\begin{multline}\tensor{\bm{G}}_s(\bm{r},\bm{r}')=\frac{ik_0^3}{8\pi^2\epsilon_0}e^{2ik_0R_0}\int_{\theta_0}^{\frac{\pi}{2}}\int_{0}^{2\pi}\tensor{\mathcal{\bm{F}}}(\theta',\varphi')\exp\bigg(-ik_0\Big[(x+x')\sin\theta'\cos\varphi'+\\(y+y')\sin\theta'\sin\varphi'+(z+z')\cos\theta'\Big]\bigg)\sin\theta' d\theta' d\varphi'\end{multline}
where
\begin{equation}\tensor{\mathcal{\bm{F}}}(\theta',\varphi')=\begin{bmatrix}\cos^2\theta'\cos^2\varphi'+\sin^2 \varphi'&-\sin^2\theta'\sin\varphi'\cos\varphi'&-\sin\theta'\cos\theta'\cos\varphi'\\
-\sin^2\theta'\sin\varphi'\cos\varphi'&\cos^2\theta'\sin^2\varphi'+\cos^2\varphi'&-\sin\theta'\cos\theta'\sin\varphi'\\
-\sin\theta'\cos\theta'\cos\varphi'&-\sin\theta'\cos\theta'\sin\varphi'&\sin^2\theta'\end{bmatrix}\end{equation}
\par
\begin{figure}[h!]
\includegraphics[width=0.4\textwidth]{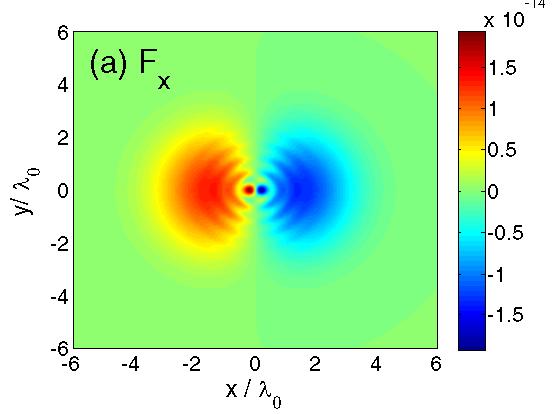}
\includegraphics[width=0.4\textwidth]{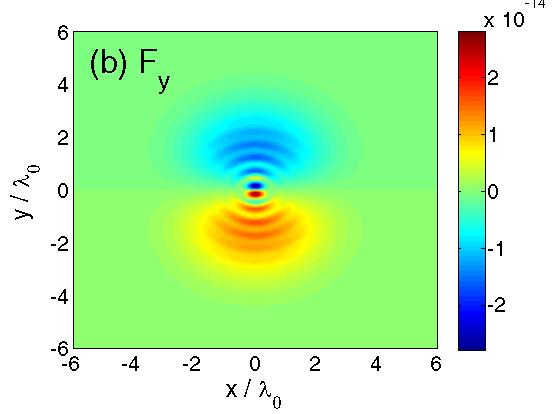}
\includegraphics[width=0.4\textwidth]{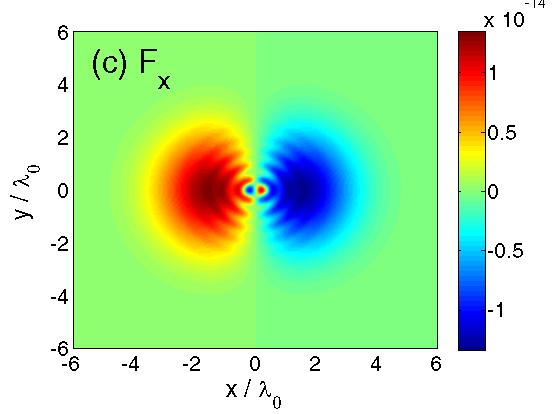}
\includegraphics[width=0.4\textwidth]{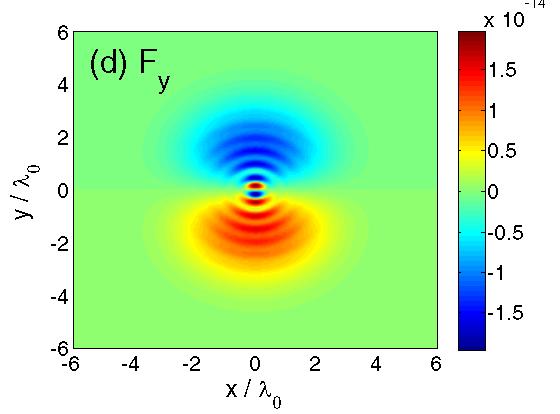}
\includegraphics[width=0.4\textwidth]{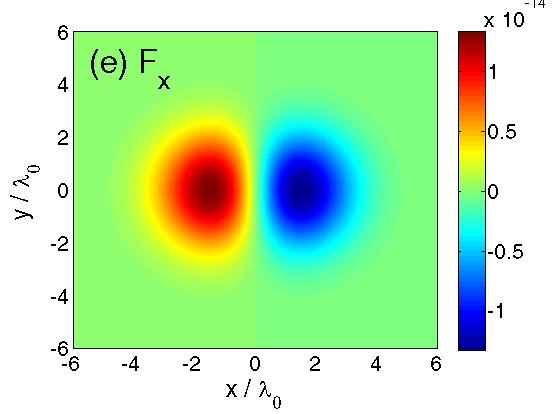}
\includegraphics[width=0.4\textwidth]{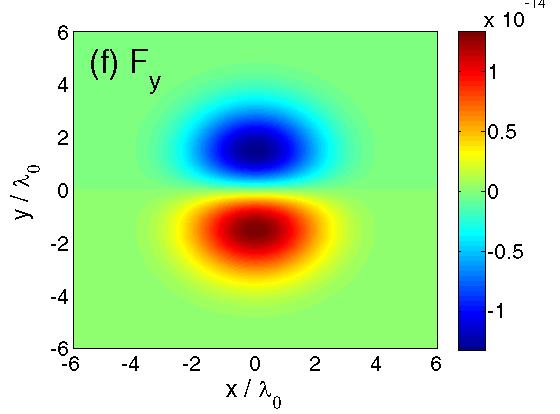}
\caption{\label{Fxy}Optical force exerted upon a 110 nm nanosphere with refractive index 2.5 at the $z=0$ plane. (a),(b) For   $R_0=(2n-1)\frac{\lambda_0}{4}$. (c),(d) For $R_0=(2n+1)\frac{\lambda_0}{4}$. (e),(f) In the absence of the spherical mirror. These results are obtained for $I_0=2[\frac{mW}{\mu m^2}]$ and $w_0=3\lambda_0$}
\end{figure}
\begin{figure}[h!]
\begin{tabular}{cc}
$\mathbf{R_0=(2n-1)\frac{\lambda_0}{4}}$&$\mathbf{R_0=(2n+1)\frac{\lambda_0}{4}}$\\
\includegraphics[width=0.4\textwidth]{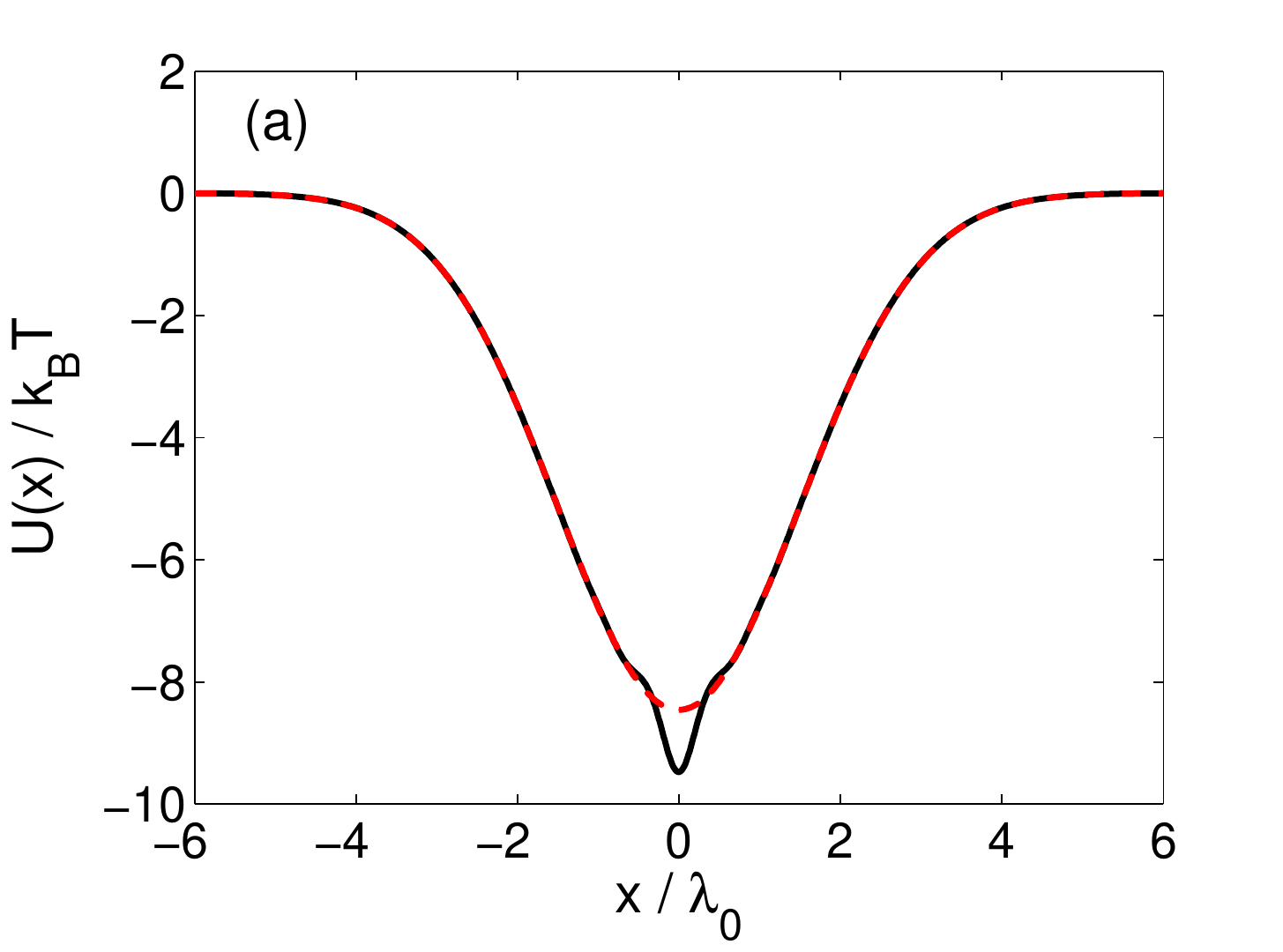}&
\includegraphics[width=0.4\textwidth]{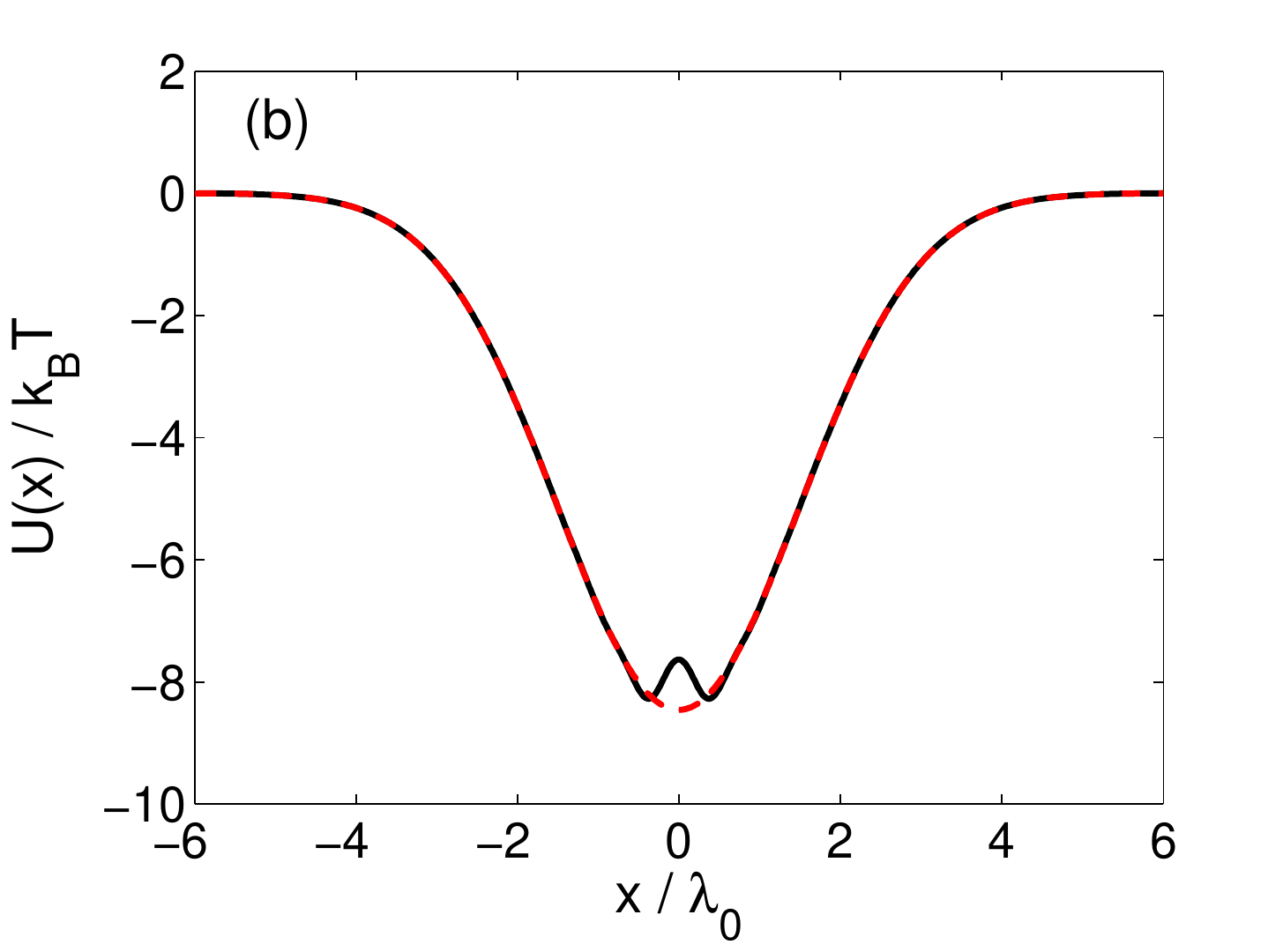}\\
\includegraphics[width=0.4\textwidth]{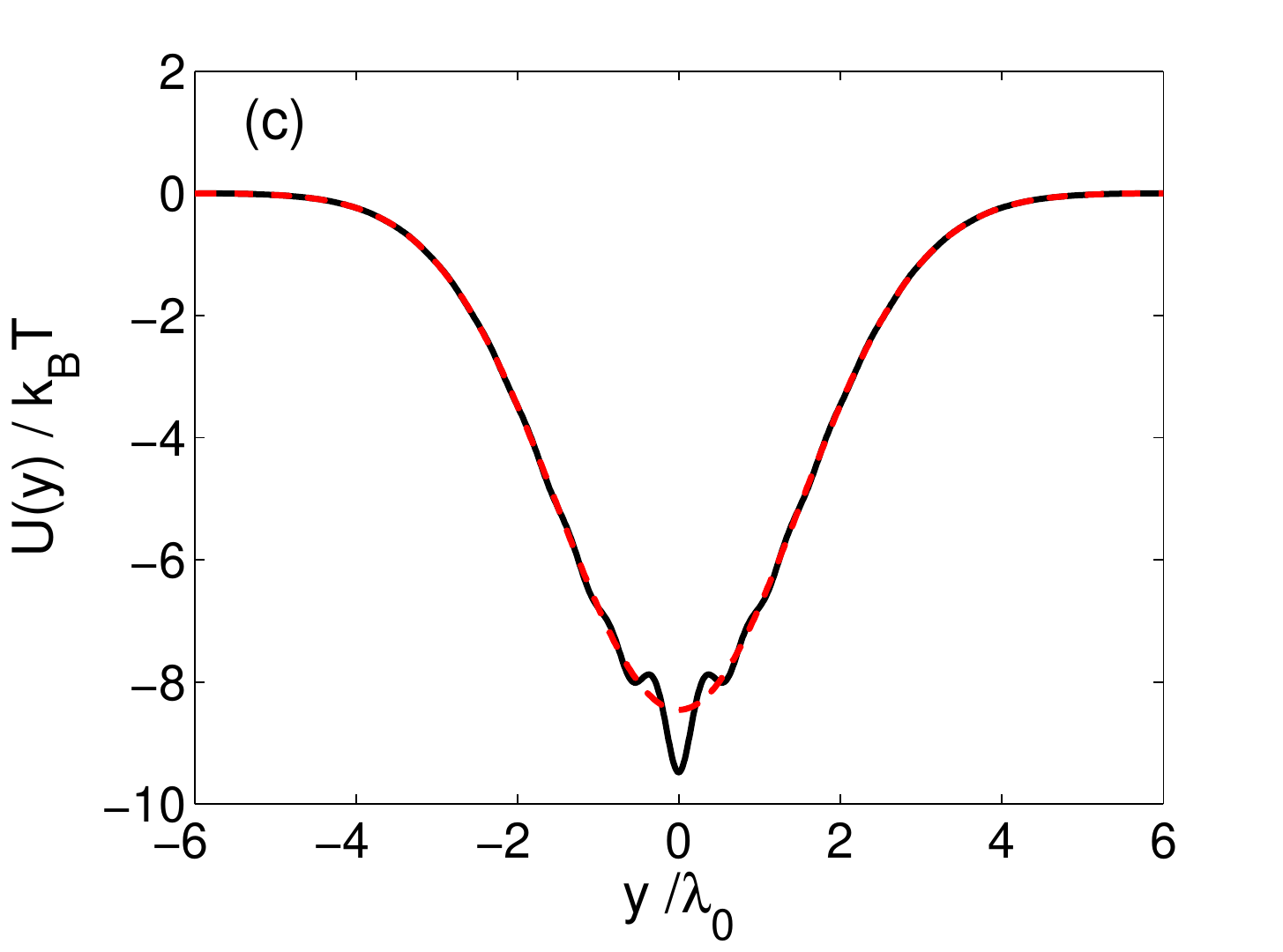}&
\includegraphics[width=0.4\textwidth]{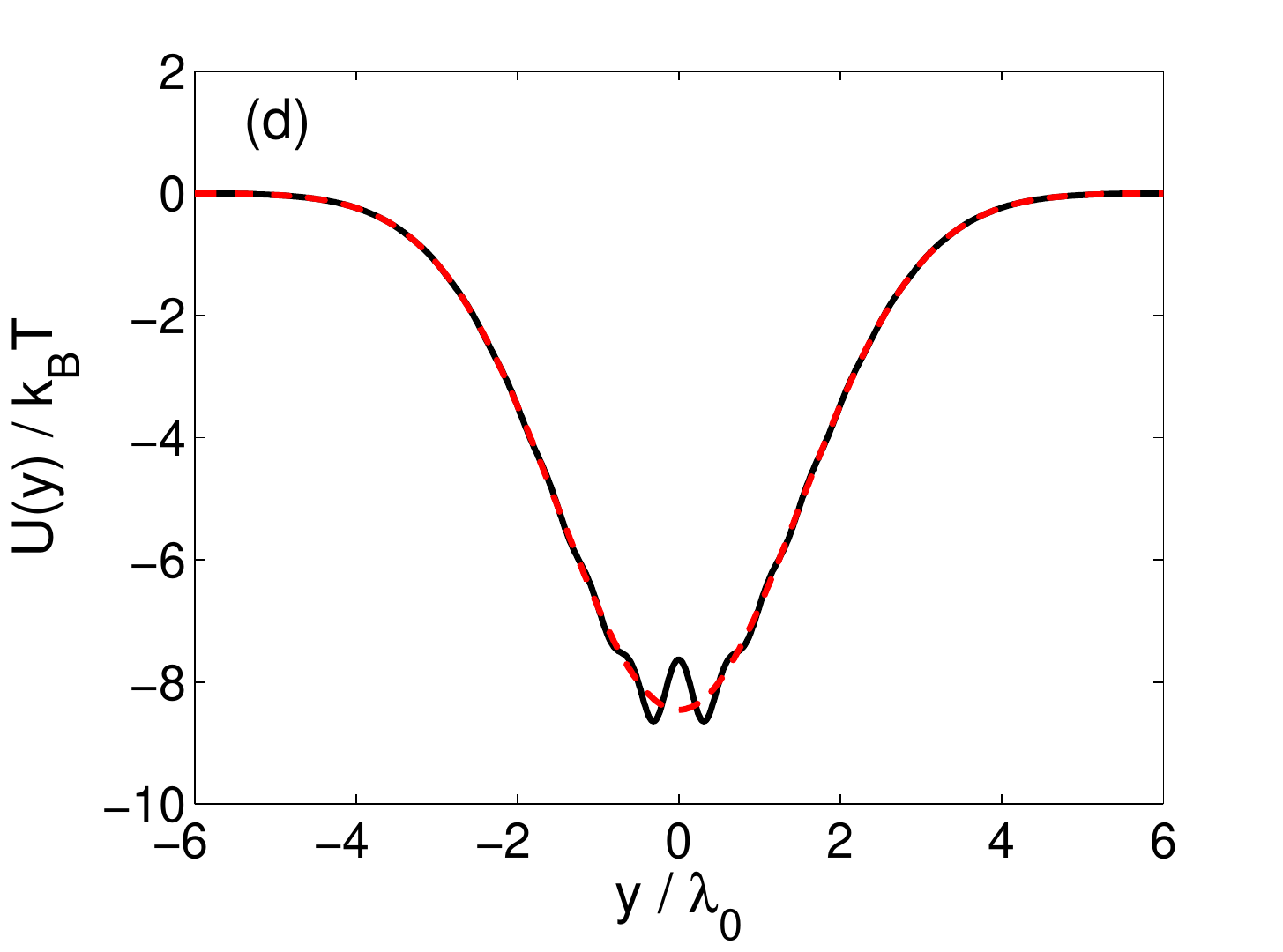}
\end{tabular}
\caption{\label{U}The trapping potential. (a) Along the $x$-axis for $R_0=(2n-1)\frac{\lambda_0}{4}$. (b) Along the $x$-axis for $R_0=(2n+1)\frac{\lambda_0}{4}$. (c) Along the $y$-axis for $R_0=(2n-1)\frac{\lambda_0}{4}$. (d) Along the $y$-axis for $R_0=(2n+1)\frac{\lambda_0}{4}$. The red-dashed lines show the backaction free results obtained in the absence of the spherical mirror}
\end{figure}

The force exerted upon a 110-nm-radius nanosphere with refractive index 2.5 by an optical tweezer when $I_0=2[\frac{mW}{\mu m^2}]$, $w_0=3\lambda_0$, and $\lambda_0=1064 [nm]$ wavelength is numerically calculated and plotted in Fig. \ref{Fxy}. Furthermore, the waist radius of the optical tweezer is considered $w_0=3\lambda_0$. Note that the calculation is carried out at two different radii $R_0=(2n-1)\frac{\lambda_0}{4}$, and $R_0=(2n+1)\frac{\lambda_0}{4}$, and the results are given in Figs. 2a,2b and Figs. 2c,2d, respectively. These figures clearly demonstrate that the results significantly depend on the radius of the spherical mirror. Furthermore, the impact of the backaction on the exerted force can be weighed by comparing the results in Figs. 2a-2d with the results obtained in the absence of the spherical mirror which are shown in Figs. 2e-2f. According to these figures, the backaction increases the maximum force, and modifies the force profile. Looking into Figs. 2c,2d shows that when $R_0=(2n+1)\frac{\lambda_0}{4}$ the backaction shifts the trapping position of the particle from the center of the beam, and results in a dual trap. This phenomenon can be better seen in Fig. \ref{U} where the trapping potentials are depicted along the $x$ and $y$ axes. Note that our simulations show that the mirror can change the real part of the polarizability up to 10 percent while as seen in Fig. \ref{Fxy} the backaction has a greater impact on the force exerted upon the nanosphere. This difference  originates from the new force term in Eq. \ref{eq:F2} associated with the gradient of the scattering Green function of the structure.\par
\begin{figure}[h!]
\includegraphics[width=0.4\textwidth]{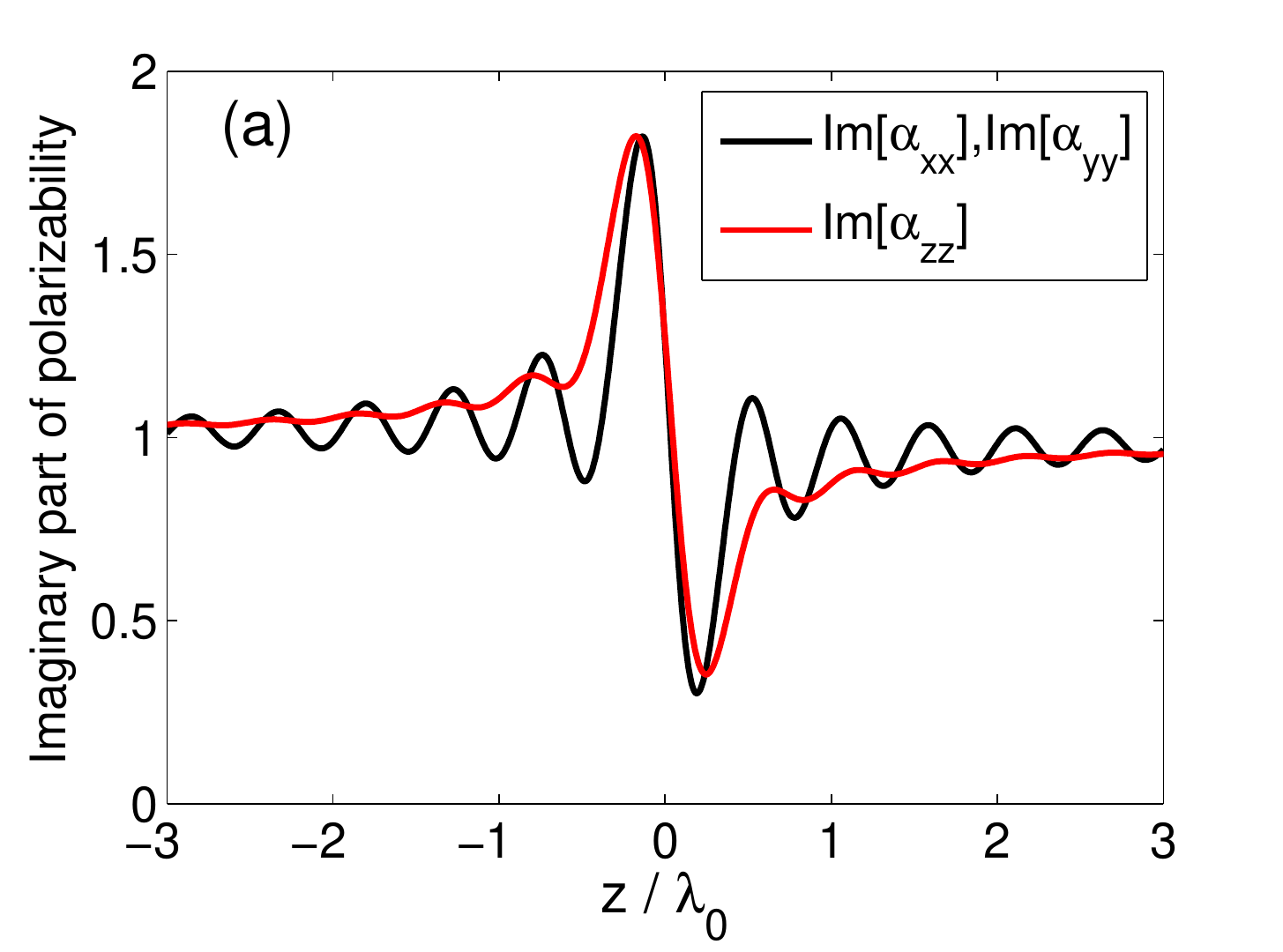}
\includegraphics[width=0.4\textwidth]{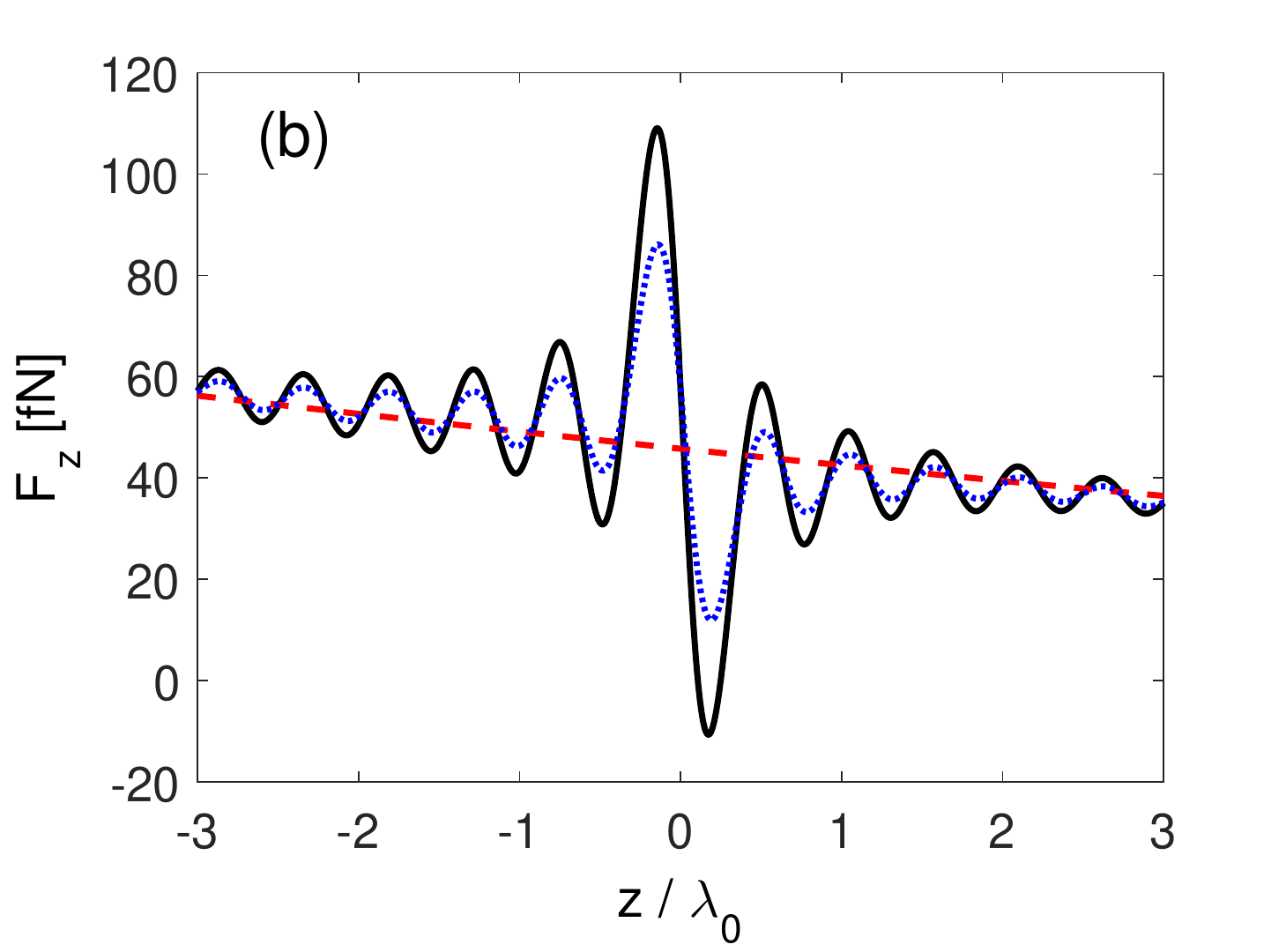}
\caption{\label{Fz}(a) The imaginary part of the polarizability normalized to the imaginary part of the free-space polarizability along the $z$-axis. (b) The force exerted upon the nanosphere along the $z$-axis. The red-dashed line is the force in the absence of the spherical mirror. The blue-dotted line is the scattering force in the presence of the nanosphere neglecting the contribution of the gradient of the scattering Green function. These results are obtained for $R_0=(2n-1)\frac{\lambda_0}{4}$}
\end{figure}

Finally, the influence of the backaction on the scattering force exerted upon the particle along the axis of the beam is studied. Fig. \ref{Fz} (a) shows the imaginary part of the polarizability of the nanosphere  along the $z$-axis. As it can be seen in this figure, the imaginary part of the polarizability is now a function of the position of the nanosphere since the strength of the backaction effect depends on how close the particle is to the center of the spherical mirror. Fig. \ref{Fz} (b) shows the force exerted upon the nanosphere along the $z$-axis. According to this figure, the backaction can counteract the scattering force at the beam waist ($z=0$), and thus form a stable three dimensional trap.\par
\subsection{\label{sec3:b}Circular nano-hole structure}
\begin{figure}[h!]
\includegraphics[width=0.4\textwidth]{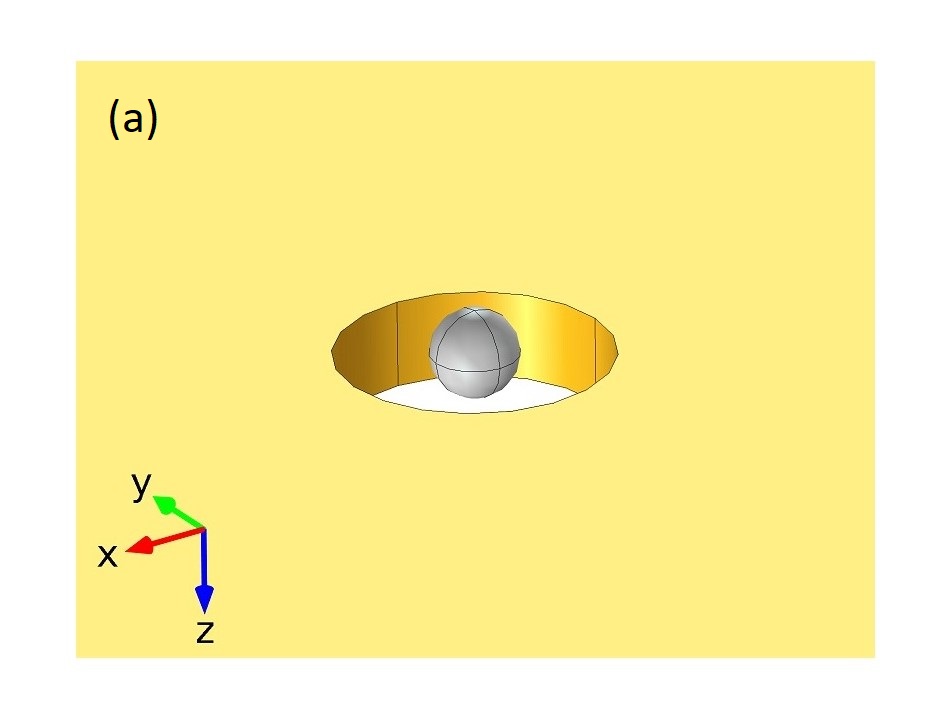}
\includegraphics[width=0.4\textwidth]{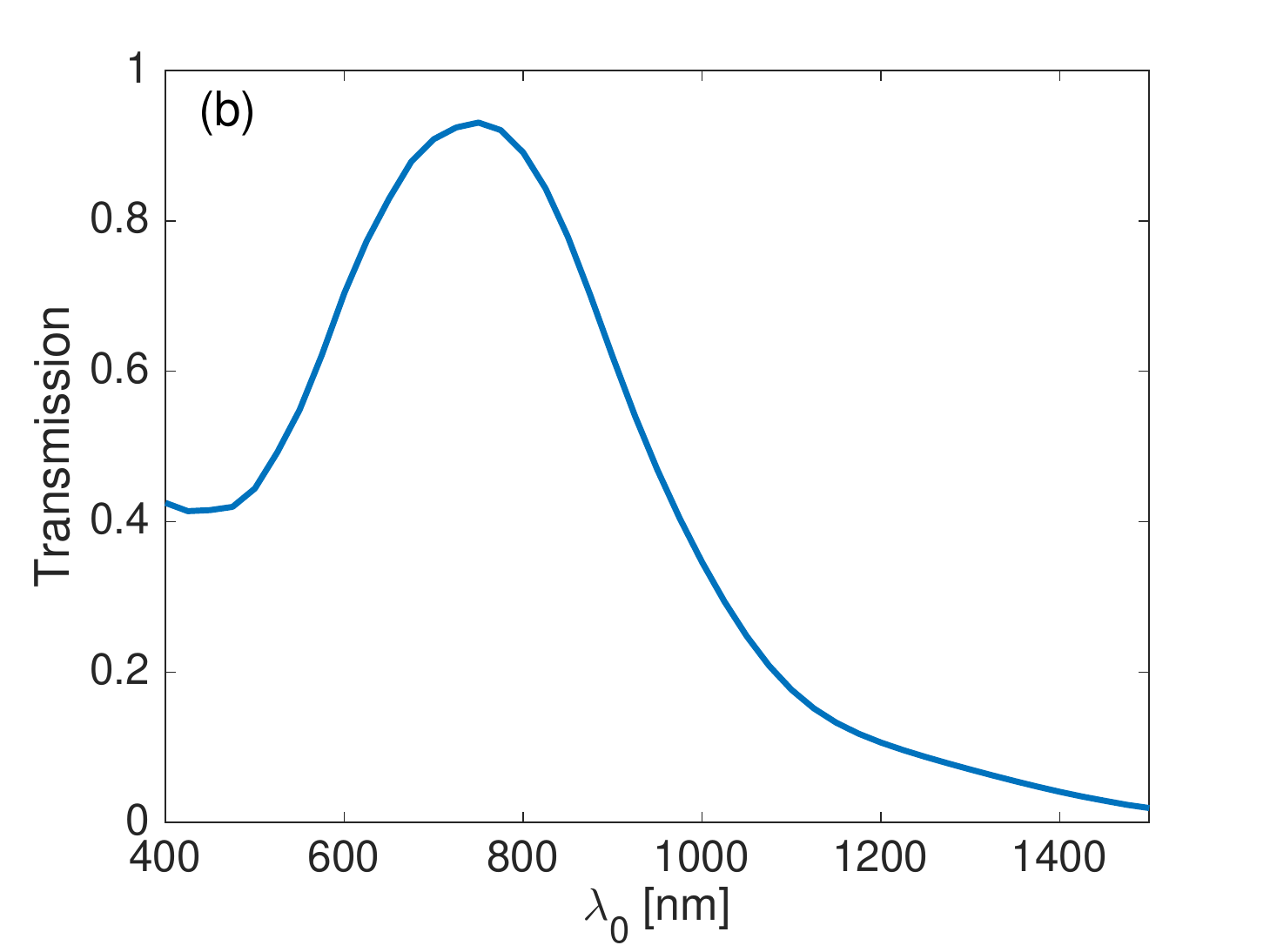}
\includegraphics[width=0.4\textwidth]{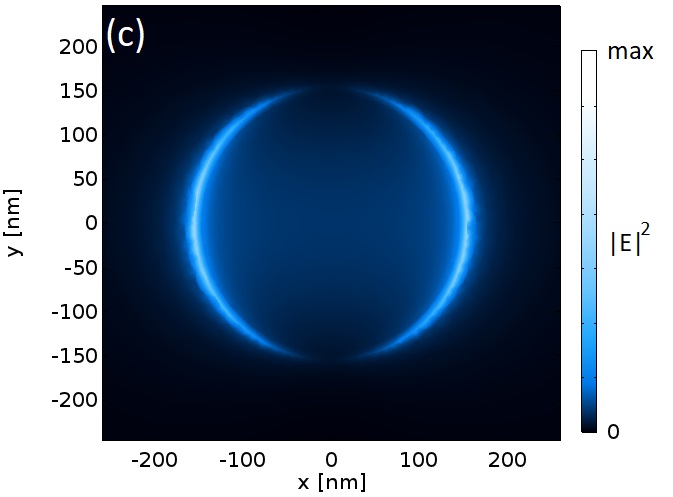}
\includegraphics[width=0.4\textwidth]{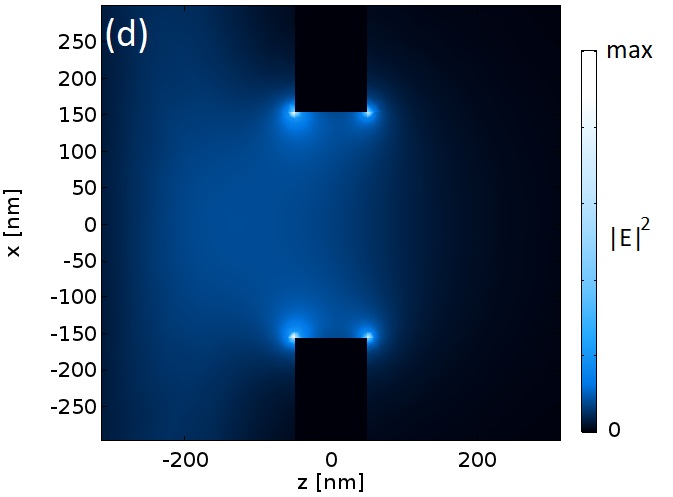}
\caption{\label{CNH_scheme}(a) Schematic of the CNH structure. (b) Transmission spectrum of CNH structure in the absence of the nanosphere. (c), (d) Intensity profiles of the electromagnetic fields in the absence of the nanosphere.}
\end{figure}

Here, we investigate the trapping of a Rayleigh nanosphere in a structure with a circular nano-hole. The schematic of this configuration is depicted in Fig. \ref{CNH_scheme}(a). As shown in this figure, a 310nm-diameter circular hole is drilled inside a 100nm-thick gold film. An $x$-polarized Gaussian beam is incident upon the aperture which is utilized for trapping of a 50nm-radius polystyrene nanosphere. The transmission spectrum of this structure is shown in Fig. \ref{CNH_scheme}(b). According to this figure, it has a pick at 750nm wavelength. However, since the quality factor of this resonance is very low ($Q\sim 1.5$), this structure cannot be analyzed using a single resonance mode of the structure. The wavelength of the incident field is adjusted at 800nm which, as in Ref. \cite{juan2009self}, is slightly above the resonance wavelength of the structure. Furthermore, the intensity of the electromagnetic fields supported by this structure in the absence of the nanosphere is depicted in Figs. \ref{CNH_scheme}(c),(d).\par
\begin{figure}[h!]
\includegraphics[width=0.4\textwidth]{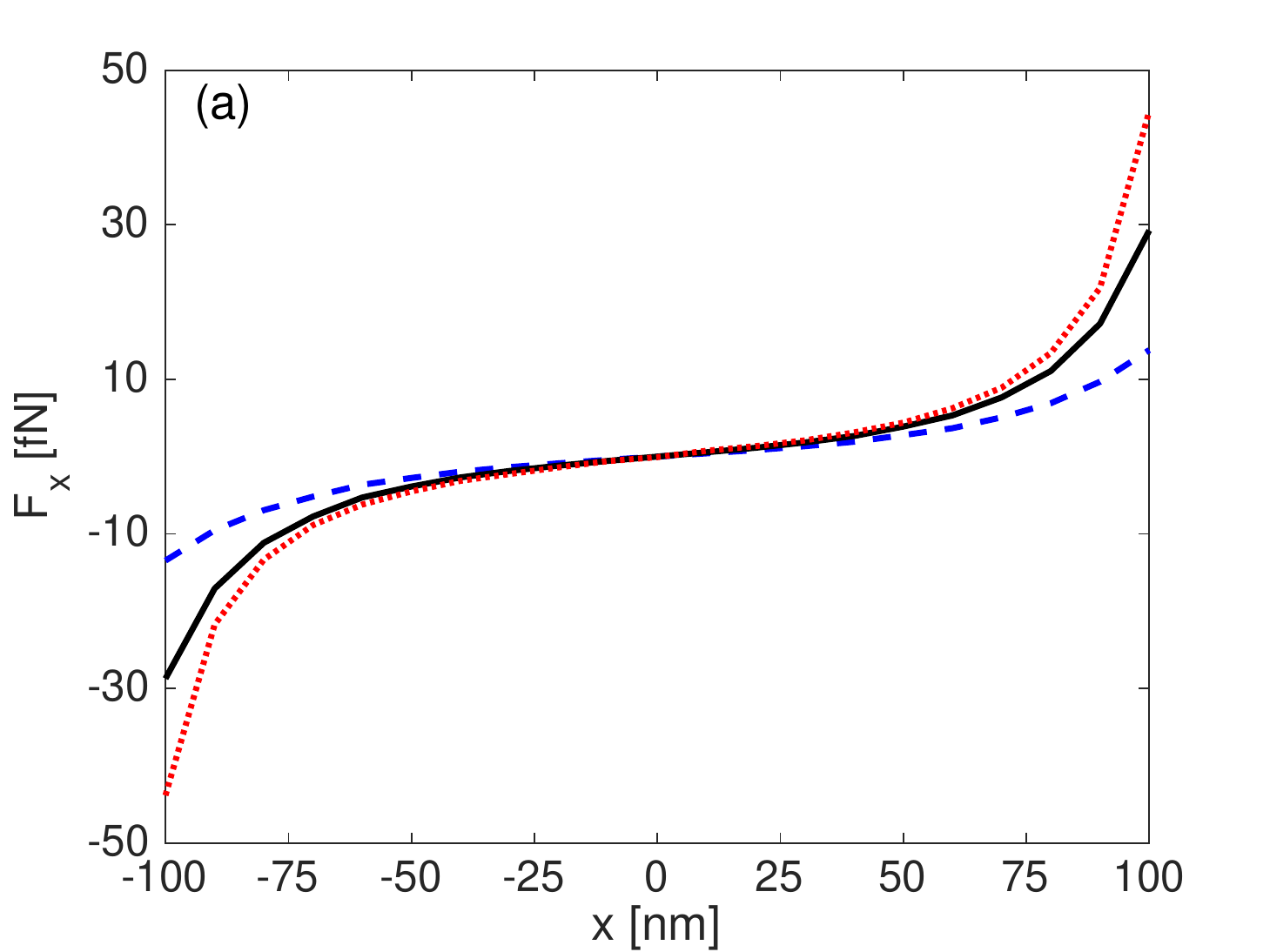}
\includegraphics[width=0.4\textwidth]{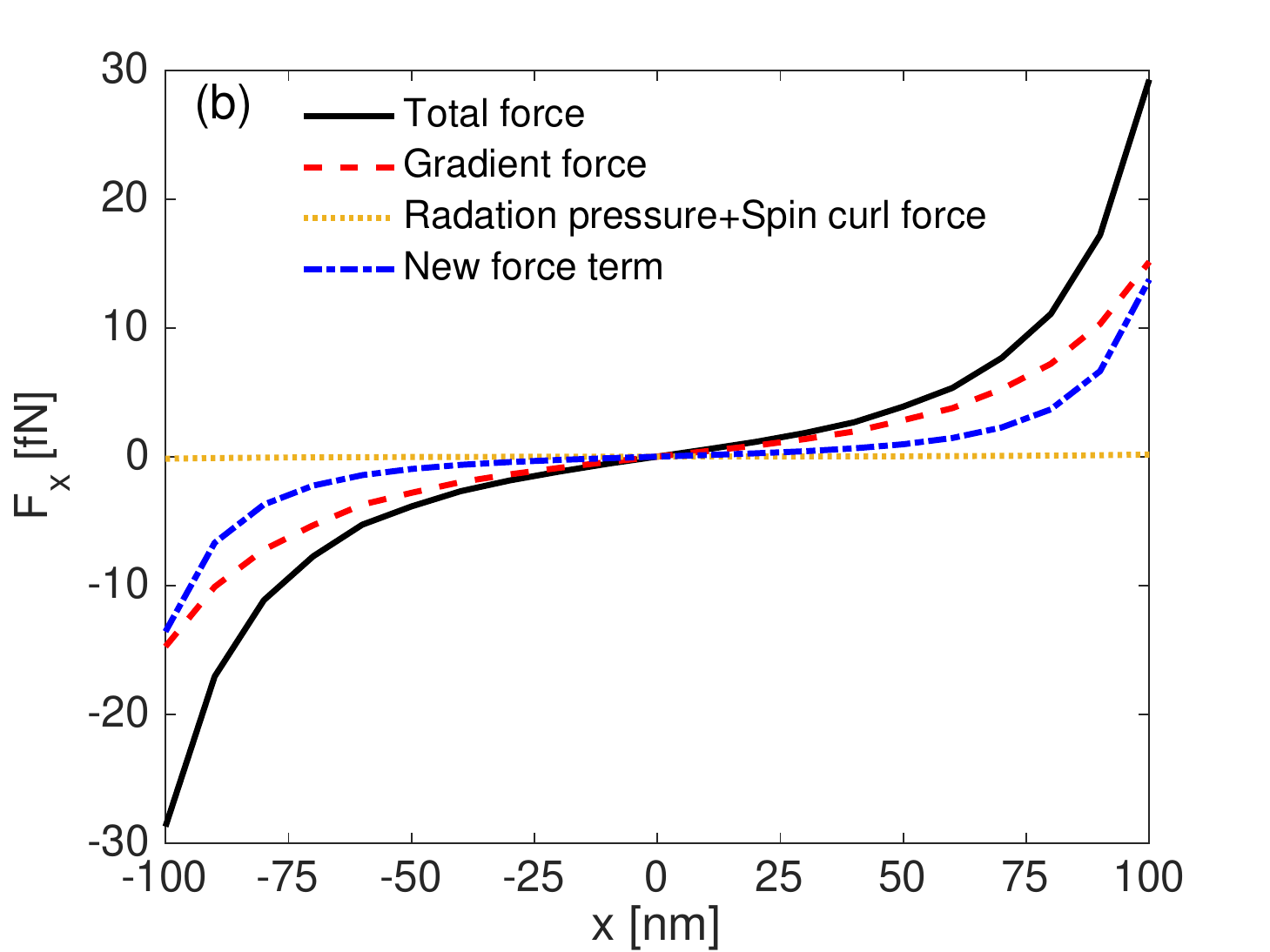}
\includegraphics[width=0.4\textwidth]{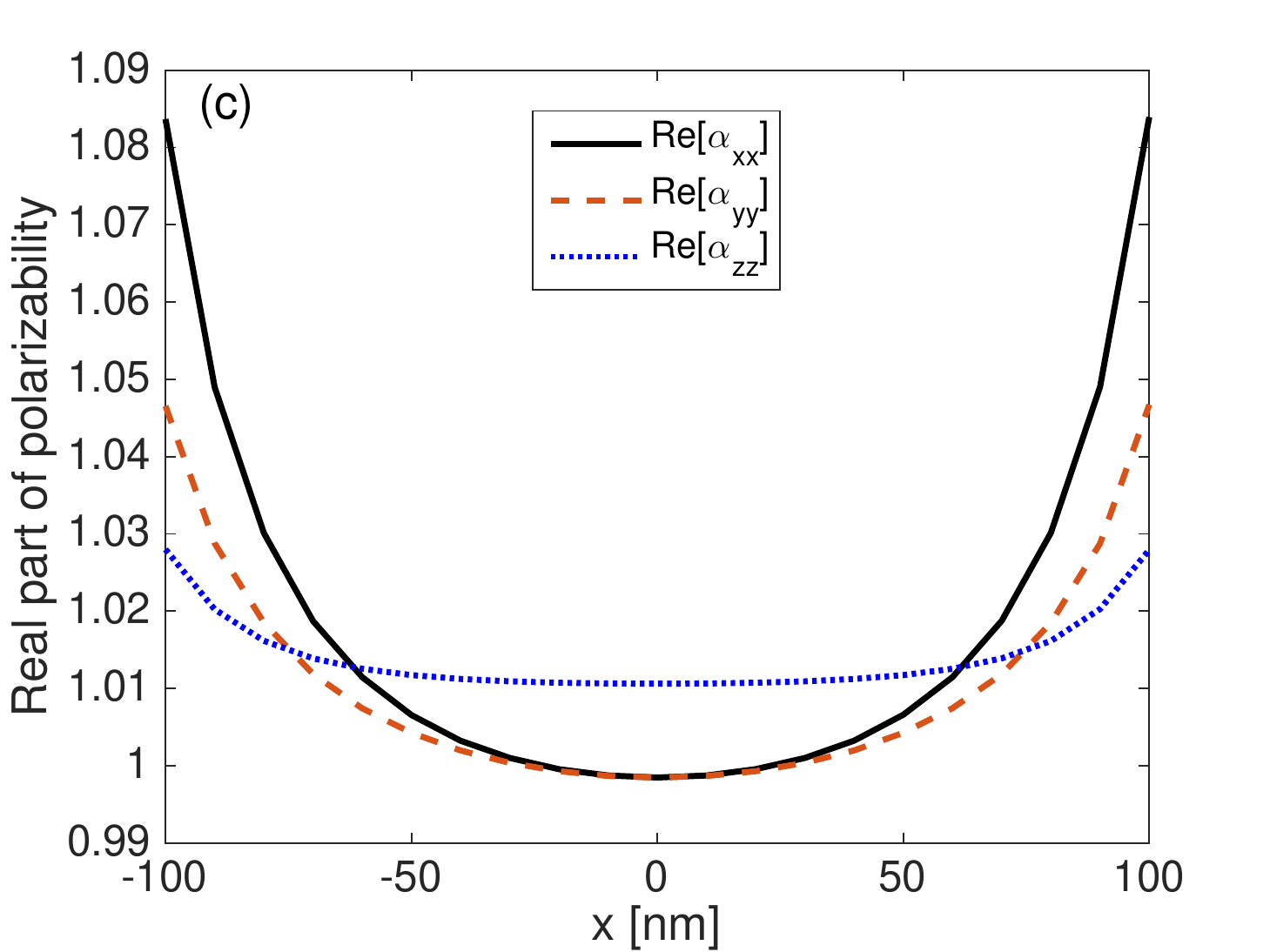}
\caption{\label{CNH_Fx}(a) Optical force exerted upon a 50nm-radius polystyrene nanoparticle along the $x$-direction at the entry of the aperture. The force is calculated with three different methods: Maxwell stress tensor(red-dotted), dipole approximation including the backaction effect (solid black), and conventional dipole approximation without considering the backaction effect (blue-dashed). (b) Contribution of different force terms. (c) Real part of the polarizability of the nanosphere normalized to the polarizability of the nanosphere in the free space.}
\end{figure}
\begin{figure}[h!]
\includegraphics[width=0.4\textwidth]{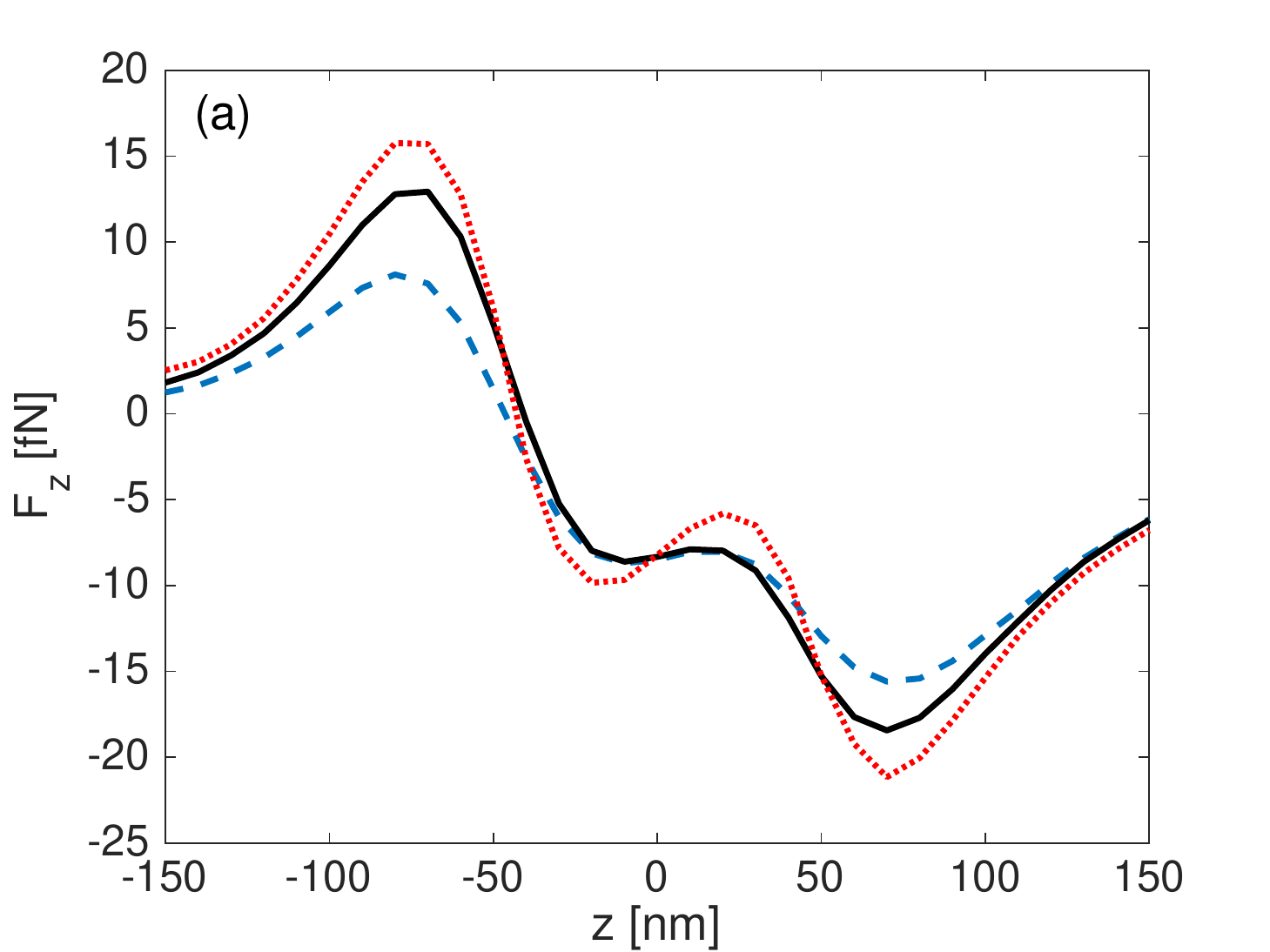}
\includegraphics[width=0.4\textwidth]{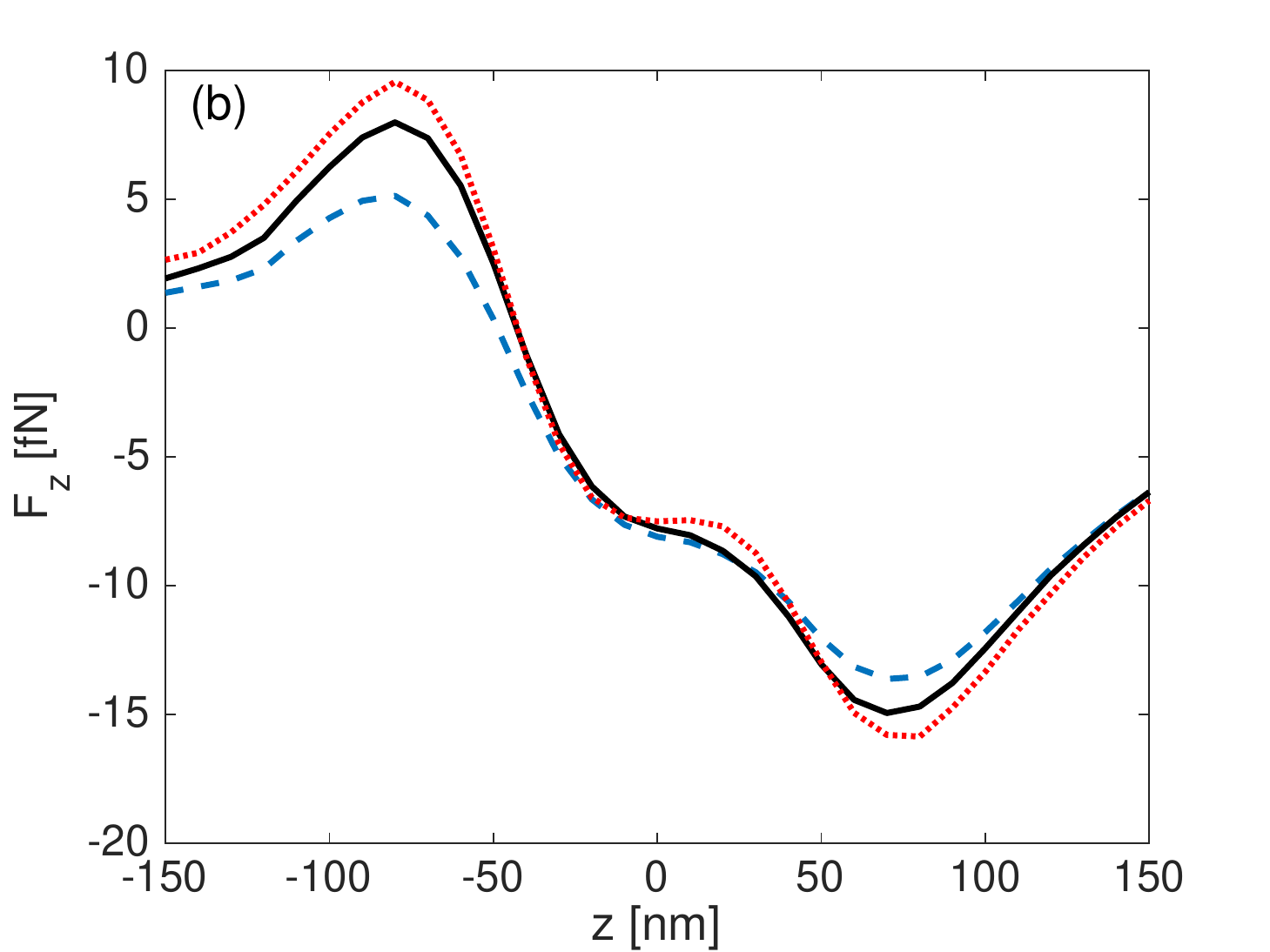}
\caption{\label{CNH_Fz}  Optical force exerted upon a 50nm-radius polystyrene nanoparticle along $z$-direction at (a) x=100nm (b)  x=90nm from the center of the hole. (Red-dotted) Maxwell stress tensor, (solid black) dipole approximation including the backaction effect, (blue-dashed) conventional dipole approximation without considering the backaction effect.}
\end{figure}

Fig. \ref{CNH_Fx}(a) shows the calculated force along the $x$-direction at the entry of the aperture. In this figure, in addition to the force calculated from our formalism, the results of the force calculation by applying the Maxwell stress tensor, and the conventional dipole approximation neglecting the impact of the scattering Green function (referred to as the perturbative method in Ref. \cite{juan2009self}), are also depicted. This figure definitely helps us to distinguish between the contribution of the scattering Green function referred to as the backaction effect and the contribution of the higher-order multipoles in the CNH structure. The difference between the solid black line and the blue-dashed line in Fig. \ref{CNH_Fx}(a) is due to the backaction effect, and the difference between the solid black line and red-dotted one is due to the contribution of the higher-order multipoles. This figure clearly indicates that the proposed formalism can accurately predict the force in the CNH structure, and the contribution of higher multipole orders becomes important only when the particle is almost touching to the walls of the aperture. Fig. \ref{CNH_Fx}(b) discribes the contribution of different force terms introduced in Eq. \ref{eq:F2}. Furthermore, the real part of the polarizability of the particle is depicted in Fig. \ref{CNH_Fx}(c). From these figures, it can be figured out that the backaction cannot significantly affect the real part of the polarizability of the particle, and the backaction effect seen in CNH structure mostly stems from  the new force term introduced in Eq. \ref{eq:F2}. Eventually, we also calculate the force exerted upon the nanosphere along the $z$-direction which is depicted in Fig. \ref{CNH_Fz}.\par
\section{\label{sec4}Conclusion}
In this paper, a closed form expression is given for the exerted force upon a Rayleigh particle in nonfree space by proposing  a mathematical formalism based on the scattering Green function of the structure wherein the Rayleigh particle is put. In this manner, the dipole approximation for the calculation of the total force is modified to include the backaction effect. It was shown that the backaction fields due to the presence of scattering objects can modify gradient force, radiation pressure, and spin curl force by changing the polarizability of the particle. Furthermore, it was shown that the backaction brings about a new force term which depends on the gradient of the scattering Green function of the structure.\par
The proposed formalism is a beneficial tool for studying trapping of Rayleigh nanoparticles in nanostructures, and can facilitate the design of nanometric optical tweezers which are to benefit from the backaction effect. It is also worth noting that  modeling the backaction effect by the scattering Green function of the structure can be employed to analyze backaction cooling of nanoparticles in nonresonant structures.

\appendix*
\section{\label{sec:Hamiltonian}Hamiltonian formalism in resonant structures}
Assume a Rayleigh nanosphere is trapped inside a resonant structure with resonance frequency $\omega_c$. The electric field of the resonant mode is considered as
\begin{equation}\bm{E}(\bm{r})=i\sqrt{\frac{2\hbar \omega_c}{\epsilon_0 V_m}} a \bm{u}(\bm{r})\end{equation}
where $a$, and $V_m$ are the mode amplitiude, and mode volume of the resonant mode, respectively, and $\bm{u}(\bm{r})$ is the normalized field profile of the resonant mode. The interaction between the nanosphere and the optical mode can be described with the interaction part of Hamiltonian which is given by
\begin{equation}H_{int}=-\frac{1}{4}\alpha_0|\bm{E}|^2=-\hbar Aa^\dagger a \Psi(\bm{r}_p)\end{equation}
where $\Psi(\bm{r}_p)=|\bm{u}(\bm{r}_p)|^2$ is the normalized intensity profile of the resonant mode, and $A=\frac{\omega_c\alpha_0}{2\epsilon_0V_m}$ is the maximum resonance frequency shift of the resonant mode due to the presence of the nanosphere.\par
The equation of motion for $a$ can be written as
\begin{equation}\frac{da}{dt}=i\Delta a - \frac{\kappa}{2}a+i A \Psi(\bm{r}_p)a+\sqrt{\kappa_{ex}}\Omega\end{equation}
where $\Delta=\omega_L-\omega_c$ is the detuning of the laser frequency, $\omega_L$, from the resonance frequency of the cavity. $\kappa$, and $\kappa_{ex}$ are the total decay rate and external decay rate of the cavity, respectively, and $\Omega$ is the driving strength of the laser. The steady state value of $a$ is given by 
\begin{equation}a=\frac{\sqrt{\kappa_{ex}}\Omega}{\frac{\kappa}{2}-i(\Delta+A\Psi(\bm{r_p}))}\end{equation}
which clearly depends on the position of the nanosphere. It should be noted that, similar to Eq. \ref{eq:E_sep}, $a$ can also be separated to two parts. The first part is the mode amplitude in the absence of the nanosphere, and is given by
\begin{equation}a_0=\frac{\sqrt{\kappa_{ex}}\Omega}{\frac{\kappa}{2}-i\Delta}\end{equation}
The second part, is the backaction mode amplitude which can be obtained from
\begin{equation}a_{ba}=\frac{iA\Psi(\bm{r}_p)}{\frac{\kappa}{2}-i(\Delta+A\Psi(\bm{r}_p))}a_0\end{equation}
Now, it can be easily shown that the equivalent dipole moment of the nanosphere can be written as
\begin{equation}\bm{p}=\alpha_0 \bm{E}(\bm{r}_p)=\frac{\alpha_0}{1-\frac{iA}{\frac{\kappa}{2}-i\Delta}\Psi(\bm{r}_p)}\bm{E}_0(\bm{r}_p)\end{equation}
and thereby the polarizability of the nanosphere is given by
\begin{equation}\alpha=\frac{\alpha_0}{1-\frac{iA}{\frac{\kappa}{2}-i\Delta}\Psi(\bm{r}_p)}\end{equation}


\bibliography{reference}

\end{document}